\documentclass[epj]{svjour}
\usepackage{amsfonts}
\usepackage{amsmath}
\usepackage{amssymb}
\usepackage{graphicx}

\begin{document}

\title{Magnetized color flavor locked state and compact stars}

\author{R. Gonz\'{a}lez Felipe\inst{1,2} \and D Manreza Paret\inst{3} \and A. P\'{e}rez Mart\'{\i}nez\inst{4}}

\institute{Instituto Superior de Engenharia de Lisboa, Rua Conselheiro Em\'{\i}dio Navarro, 1959-007 Lisboa, Portugal \and Centro de F\'{\i}sica Te\'{o}rica de Part\'{\i}culas, Instituto Superior T\'{e}cnico, Avenida Rovisco Pais, 1049-001 Lisboa, Portugal, \email{gonzalez@cftp.ist.utl.pt} \and Universidad de la Habana, Facultad de F\'{\i}sica, San L\'{a}zaro y L, La Habana, 10400, Cuba, \email{dmanreza@fisica.uh.cu} \and Instituto de Cibern\'{e}tica, Matem\'{a}tica y F\'{\i}sica (ICIMAF), Calle E esq 15 No. 309 Vedado, La Habana, 10400, Cuba, \email{aurora@icmf.inf.cu}}

\date{ }

\abstract{The stability of the color flavor locked phase in the presence of a strong magnetic field is investigated within the phenomenological MIT bag model, taking into account the variation of the strange quark mass, the baryon density, the magnetic field, as well as the bag and gap parameters. It is found that the minimum value of the energy per baryon in a color flavor locked state at vanishing pressure is lower than the corresponding one for unpaired magnetized strange quark matter and, as the magnetic field increases, the energy per baryon decreases. This implies that magnetized color flavor locked matter is more stable and could become the ground state inside neutron stars. The mass-radius relation for such stars is also studied.}

\PACS{26.60.-c, 21.65.Qr, 26.60.Kp, 04.40.Dg}

\maketitle

\section{Introduction}
\label{sec1}

The internal composition of neutron stars as well as the  real nature of the ground state of matter moves through interconnected avenues. Bodmer~\cite{Bodmer:1971we} and Witten ~\cite{Witten:1984rs} suggested that strange quark matter (SQM) could be a stable phase of nuclear matter. This exciting result continues being a conjecture because presently it is impossible to perform laboratory experiments that confirm it. Nevertheless, this issue has attracted great attention in the astrophysical context and many works have been devoted to study the properties of the equation of state (EoS) of SQM and its connection with strange star or neutron star observables. In particular, the existence of a more compact form of matter could be a plausible explanation for the still unexplained observation of sources of gamma-$\gamma$ rays bursts. Furthermore, studies of the superconductor phases of the QCD suggest that the ground state of matter could be a superconductor phase, being a compact object the natural scenario of this phase transition.

The pioneer studies of the pairing interaction of the dense matter appeared more than thirty years ago~\cite{Barrois:1977xd,Bailin:1983bm}. Under certain conditions, SQM could undergo a phase transition to a superconductor phase, in particular at high densities and low temperature. The most symmetric phase among these phases is the so-called color flavor locked (CFL) state~ \cite{Alford:1998mk,Schafer:1999fe,Shovkovy:1999mr,Alford:2001zr,Rajagopal:2000ff,Alford:2002kj,Alford:2007xm,Alford:2004pf}.

On the other hand, there is not doubt that the role of the magnetic field in astrophysical scenarios is very important. Pulsars, magnetars, neutron stars, the emission of intense sources of X-rays could be associated to sources with intense magnetic fields around $10^{13}-10^{15}$~G or even higher fields~\cite{duncan,kouve}. Furthermore, the magnetic field intensity may vary significantly from the surface to the center of the source and theoretical estimates indicate that fields as high as $10^{19}$~G could be allowed~\cite{dong}. The relevance of the magnetic field in color superconductivity has been studied in refs.~\cite{Alford:1999pb,Gorbar:2000ms,Ferrer:2005vd,Ferrer:2006vw,Ferrer:2006ie,Ferrer:2007iw,Fukushima:2007fc,Noronha:2007wg,Alford:2010qf}. These papers have tackled, among other aspects, the modification of the pairing pattern by the external field, the formation of a gluon condensate at certain field strengths, and the boost of the applied field due to the back reaction of the color superconductor. These results support the idea that the magnetic field enhances color superconductivity. It has also been shown that magnetic fields in neutron stars with color superconducting cores are stable on time scales comparable with the age of the Universe~\cite{Alford:1999pb,Alford:2010qf}. Thus, seeking for pulsars which do not diminish their magnetic field as the star spins down could help to find evidences of the existence of a superconductor phase of quark matter inside compact stars.

Following this line of research, it is therefore worthwhile to study astrophysical observables, such as the mass-radius relation of quark stars, in a magnetized CFL phase. In ref.~\cite{Felipe:2008cm}, it was shown that magnetized strange quark matter (MSQM) becomes more stable than unpaired SQM. If the CFL superconductor phase is more stable and bound than unpaired SQM at finite density, one expects stable configurations of quark stars more compact than in the unpaired phase. These objects would be self-bound and their masses would scale with the radius as $M \sim R^{3}$, in contrast to neutron stars which have masses that decrease with increasing radius ($M \sim R^{-3}$) and are bound by gravity~\cite{itoh,Hansel}. Thus, self-bound stars could be consistent with small-radius compact objects~\cite{Lugones:2002va,Lugones:2002zd,Quan,Nice,Xu:2006qh}.

The aim of this paper is to study the role of the magnetic field in the CFL phase within the MIT bag model of confinement. Our intention is to show how the magnetic field can influence the stability of the phase and also its implications for the mass-radius relation generated by configurations where the deconfined matter is in the magnetized CFL phase. For the sake of simplicity, we shall assume that the gluonic contribution to the magnetic field inside the CFL phase is negligible. Due to the mixture of the photon field $A_{\mu}$ and the eighth component of the gluon field $G_{\mu}^8$, the `rotated' electromagnetic field is $\tilde{A_{\mu}}=A_{\mu}\cos \theta -G_{\mu}^8 \sin \theta$. The corresponding electromagnetic coupling constant is $\tilde{e}=e\cos\theta$, where the mixing angle $\theta$ depends on the gap structure and is given by $\cos\theta=g/\sqrt{e^2/3+g^2}$ ($g$ is the QCD coupling constant) for the CFL phase~\cite{Alford:1999pb,Gorbar:2000ms}. Since the rotated photon is massless, the magnetic field $\tilde{B}$ inside the CFL superconducting state is not screened. Moreover, in the region of interest, $e \ll g$ so that $\cos \theta \sim 1$ and one can consider that the magnetic field strengths inside and outside the CFL core are approximately equal, i.e. $\tilde{e}\tilde{B} \simeq eB$~\cite{Fukushima:2007fc}.

\begin{figure}[t]
\begin{center}
\includegraphics[width=8.0cm]{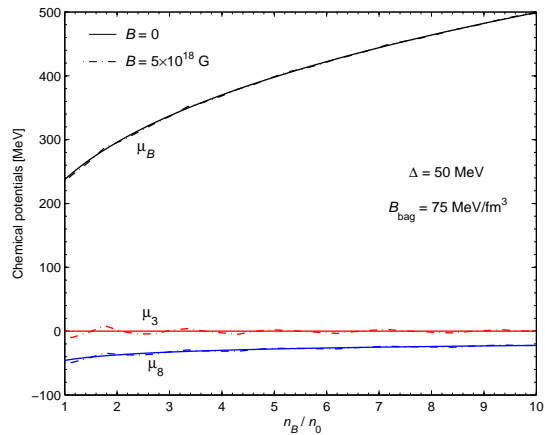}
\caption{Chemical potentials in the CFL phase for $B=0$ and $B=5\times10^{18}$~G. The curves are shown for $\Delta= 50$~MeV and $B_{\rm bag}=75$~MeV/fm$^3$.\label{chempot}}
\end{center}
\end{figure}

An important issue when studying the stability of quark matter in compact stars is the theoretical treatment of the neutrality conditions~\cite{Alford:2002kj,Buballa:2005bv}. Besides electromagnetic neutrality, color neutrality must be enforced. To guarantee the latter in the CFL phase, the chemical potentials $\mu_3$ and $\mu_8$ coupled to the color charges $T_3=\text{diag}\,(1/2,-1/2,0)$ and $T_8=\text{diag}\,(1/3,1/3,-2/3)$ should be chosen such that the $T_{3,8}$ densities vanish~\cite{Alford:2002kj}. The chemical potential for each quark $(i=u,d,s)$ is then specified by its electric and color charges, $\mu_i=\mu_{B}-Q \mu_e + T_3 \mu_3 + T_8 \mu_8$, where $\mu_B$ is the baryon chemical potential and $Q=\text{diag}\,(2/3,-1/3,-1/3)$. As it turns out, for the range of parameters considered here, to wit $eB < \mu_{B}^2$, one can show that $\mu_3 \simeq 0$ and $\mu_8 \simeq -m_s^2/(2\mu_B)$ (cf. fig.~\ref{chempot} and our discussion below eq.~\eqref{Nequality}).

In this simple framework, we shall study the behavior of the system with the variation of the relevant parameters: the bag parameter $B_{\rm bag}$, the strange quark mass $m_s$, the baryon density $n_B$, the gap parameter $\Delta$ and the magnetic field $B$. The possible mass-radius configurations of magnetized CFL stars are then obtained by solving the Tolman-Oppenheimer-Volkoff (TOV) equations. We obtain configurations of stable stars with smaller radii, which are allowed due to the compactness of matter and the presence of a strong magnetic field, since the energy per baryon at vanishing pressure is lower in this case.

The paper is organized as follows. In sect.~\ref{sec2} we briefly review the CFL phase properties in the presence of a magnetic field within the MIT bag model. In sect.~\ref{sec3}, the stability windows of CFL in the presence of a magnetic field are obtained varying the relevant input parameters of the model. Section~\ref{sec4} is devoted to the study of the mass-radius relation for CFL matter by numerically solving the TOV equations. Finally, our conclusions are given in sect.~\ref{sec5}.

\begin{figure*}[t]
      \begin{minipage}[t]{0.45\linewidth}
      \includegraphics[width=8.0cm]{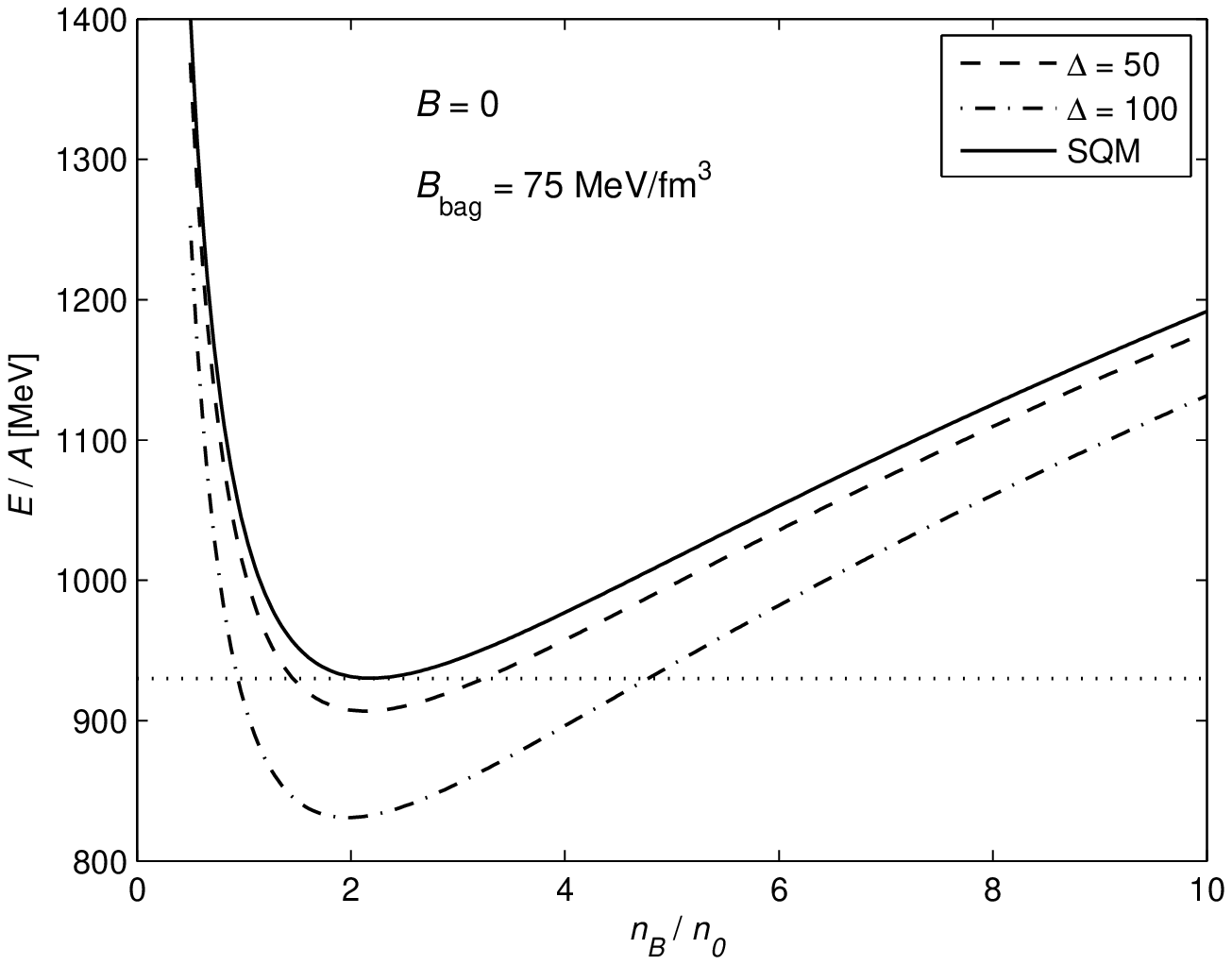}
      \end{minipage}
      \hspace{0.3cm}
      \begin{minipage}[t]{0.45\linewidth}
      \includegraphics[width=8.0cm]{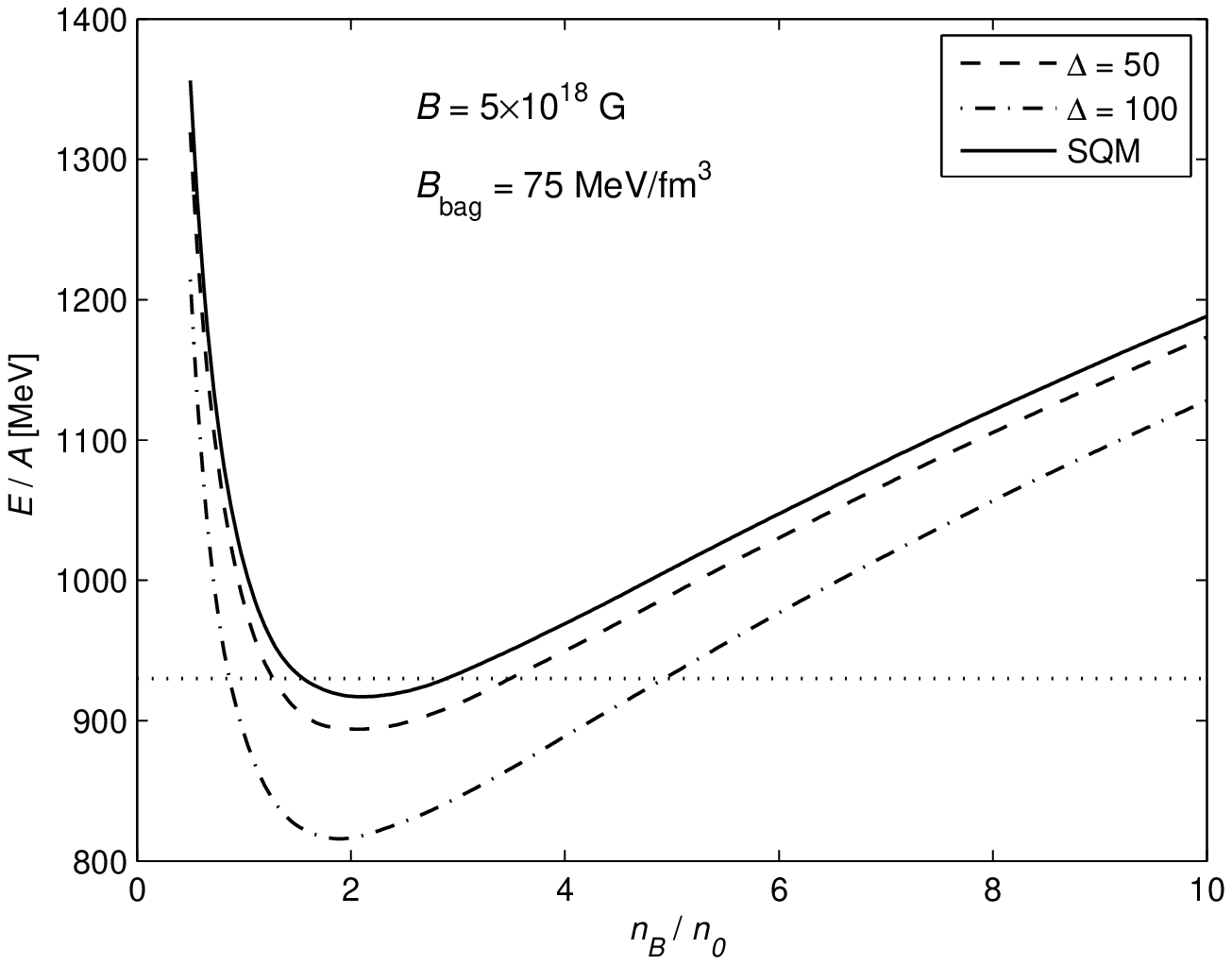}
      \end{minipage}
       \caption{The energy per baryon as a function of the baryon density for the CFL
       phase without magnetic field (left plot) and for magnetized CFL with $B=5\times
       10^{18}$~G (right plot). We take $B_{\rm bag}=75$~MeV/fm$^3$ and $\Delta=50,
       100$~MeV. For comparison, the SQM and MSQM cases are also shown. The horizontal
       dotted line corresponds to $\left. E/A\right|_{^{56}\text{Fe}} \simeq 930$~MeV.\label{EN}}
\end{figure*}

\begin{figure*}[t]
      \begin{minipage}[t]{0.45\linewidth}
      \includegraphics[width=8.0cm]{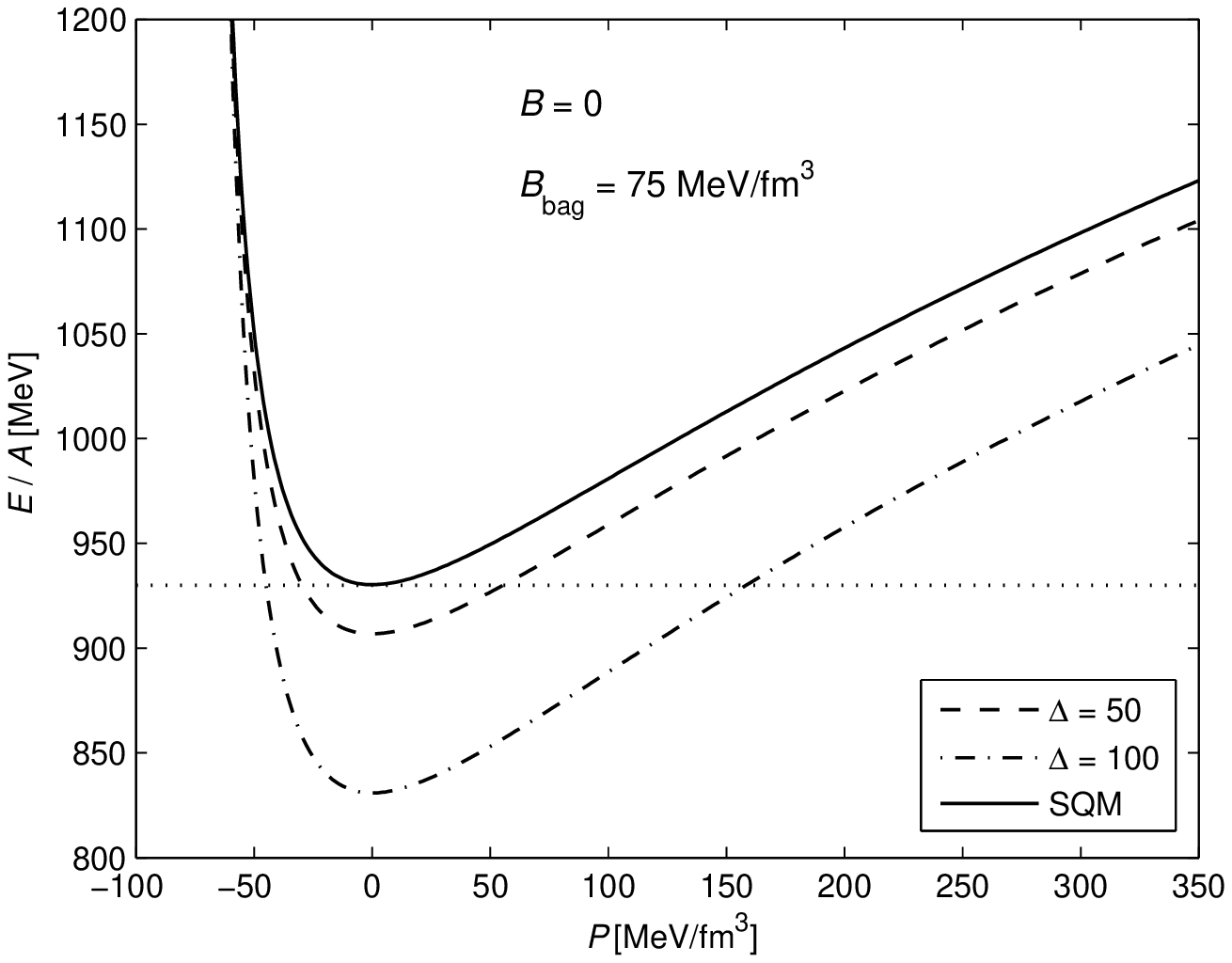}
      \end{minipage}
      \hspace{0.3cm}
      \begin{minipage}[t]{0.45\linewidth}
      \includegraphics[width=8.0cm]{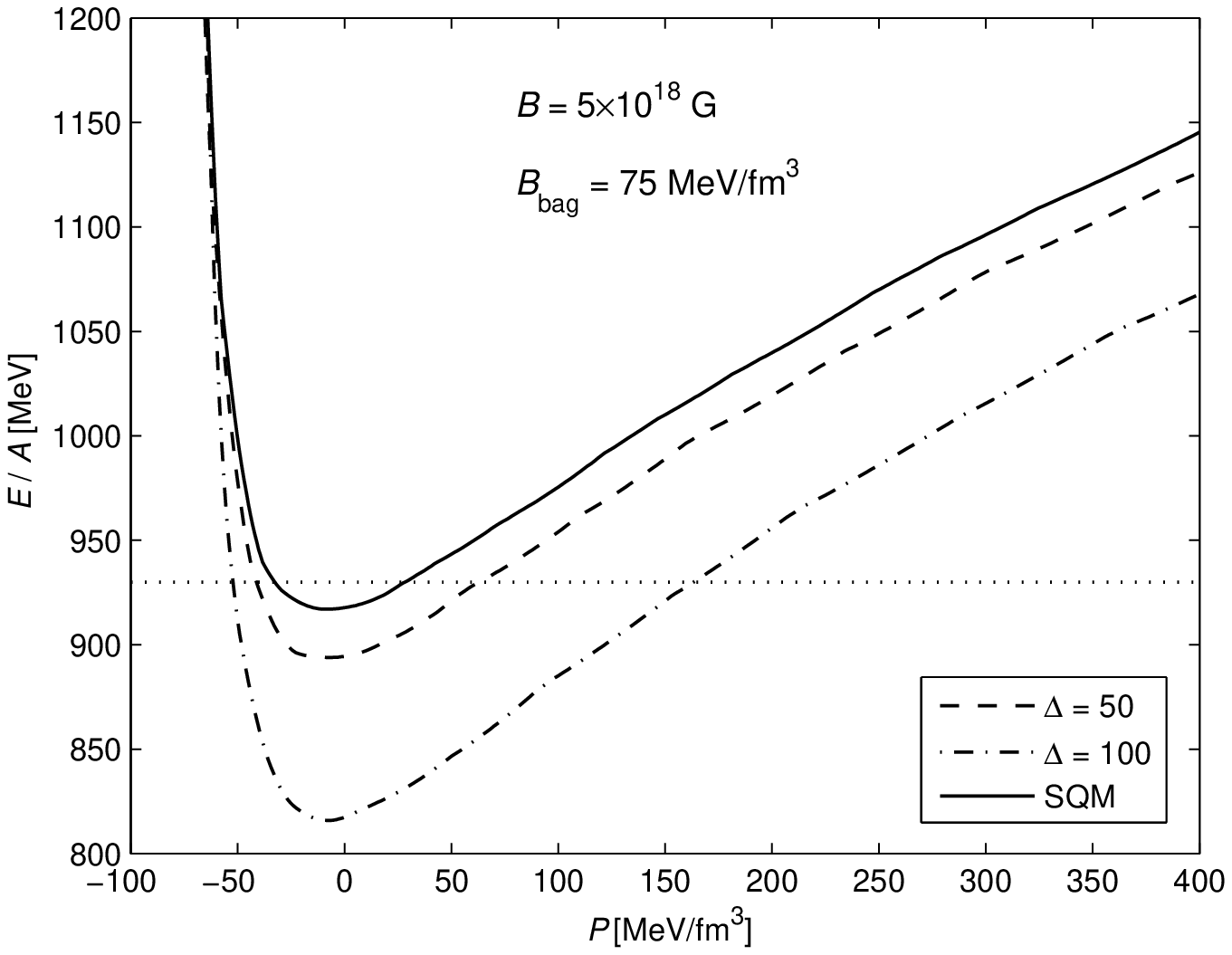}
      \end{minipage}
       \caption{The energy per baryon versus pressure for the CFL phase without magnetic field (left plot) and for magnetized CFL with $B=5\times 10^{18}$~G (right plot). We take $B_{\rm bag}=75$~MeV/fm$^3$ and $\Delta=50, 100$~MeV. The SQM and MSQM curves are also depicted. The horizontal dotted line corresponds to $\left. E/A\right|_{^{56}\text{Fe}} \simeq 930$~MeV.\label{EP}}
\end{figure*}

\section{CFL phase in the presence of a strong magnetic field}
\label{sec2}

The feasibility of the transition from SQM to CFL phase within the bag model has been studied in ref.~\cite{Lugones:2002zd}. In the present work, we shall use the same scheme to analyze the role of the magnetic field in the CFL phase and to determine if this phase continues being favored over the SQM state (unpaired phase) in the presence of a strong magnetic field. We shall not consider the corrections of the strong coupling constant which emerge from QCD. In this sense, strange quark matter is described in our framework as non-interacting. Moreover, we shall consider the limit of zero temperature since we focus on the natural scenario of quark matter in strong magnetic fields, i.e., the inner core of compact stars.

Considering a constant uniform magnetic field $B$ in the $x_3$ direction, the thermodynamical potential of MSQM (unpaired phase) can be written as~\cite{Felipe:2007vb}
\begin{align}
 \Omega_{MSQM} =  B &\sum_i {\Omega}^0_{i} \sum_{\eta=\pm 1} \sum_{n}\left(x_{i}p_{F,i}^{\eta} -
 h_{i}^{\eta\,2}\ln\frac{x_{i}
 + p_{F,i}^{\eta}}{h_{i}^\eta}\right)\nonumber\\
  &+ B_{\rm bag}, \label{pot}
 \end{align}
where the vacuum is mimicked by the bag parameter $B_{\rm bag}$, the index $i$ runs over the quark
flavors $(u,d,s)$ and electrons, and the sum in $n$ represents the sum over the Landau levels. In eq.~\eqref{pot},
\begin{align}
{\Omega}^0_{i}= \frac{d_i e_i m_{i}^2}{4\pi^2}, \quad d_e=1, d_{u,d,s}=3;
\end{align}
$x_{i}=\mu _{i}/m_{i}$ is the dimensionless chemical potential; $p_{F,i}^{\eta}$ and $h_{i}^{\eta}$ correspond to the $p_3$ component of the magnetic Fermi momentum and the magnetic mass, respectively:
\begin{eqnarray} \label{dimvar}
\begin{split}
p_{F,i}^{\eta}&=\sqrt{x_{i}^2-h_{i}^{\eta\,2}},\\
h_{i}^{\eta} &=\sqrt{\frac{B}{B^{c}_i}\, (2n + 1-\eta) +1} -\eta
y_{i}B.
\end{split}
\end{eqnarray}
In the above expressions, $B^{c}_i=m_{i}^2/|e_i|$ is the critical magnetic field and $y_i$ accounts for the anomalous magnetic moments. The sum over the Landau levels $n$ is up  to $n_{max}^{i} = I\left[\left((x_{i} + \eta y_iB)^2 -1\right)\,B^{c}_i/(2B)\right]$, where $I[z]$ denotes the integer part of $z$.

Notice that the thermodynamical potential defined in eq.~\eqref{pot} contains the contribution of Landau diamagnetism (quantization of Landau levels) as well as the Pauli paramagnetism due to the quark anomalous magnetic moments. Since the inclusion of the latter does not significantly  modify the EoS, in our analysis we shall neglect their contribution and consider only the effect of Landau diamagnetism\footnote{Including the anomalous magnetic moments just places a more restrictive upper bound on the magnetic field (beyond the saturation field $\sim 10^{18}$~G the thermodynamical quantities become complex~\cite{Felipe:2007vb}.)}.

Strictly speaking, besides the statistical contribution given in eq.~\eqref{pot}, the thermodynamical potential contains a vacuum contribution, $\Omega_{\text{vac}}$, which does not depend on the temperature and quark densities. The latter term has a field-independent ultraviolet divergence that should be renormalized. After renormalization, the following expression is obtained~\cite{Schwinger:1951nm}:
\begin{align}
\Omega_{\text{vac}}=-\sum_i \frac{(e_iB)^2}{8\pi^2}\int_0^\infty \frac{dz}{z^3} e^{-z B_i^c/B}
 \left[z \coth z-1-\frac{1}{3}z^2\right].
\end{align}
Taking, for instance\footnote{Hereafter in our calculations we shall assume these values for the quark masses.}, $m_u = m_d = 5$~MeV and $m_s = 150$~MeV and a magnetic field $B=5\times10^{18}$~G, one has $\Omega_{\text{vac}} \simeq 1.1$~MeV/fm$^{3}$. Thus, in applications to astrophysical compact objects with $eB < \mu_B^2$, the leading contribution to the EoS will come from the statistical terms of $\Omega$ given in eq.~\eqref{pot}.

The technical question about the renormalization of the gluonic contribution to the thermodynamic potential in the CFL phase is nevertheless subtle. Because there is a difference between the ordinary electromagnetism in normal quark matter and the modified one in the CFL phase (photon-gluon mixture), it remains uncertain how to remove the divergent zero-point energy in the latter case. One may hope that a substraction procedure analogous to the one employed for the electroweak gauge bosons interacting with a magnetic field could be applied. As for the renormalized gluonic contribution, one could expect that it would not significantly affect the EoS. Indeed, it only depends on the magnetic field and the photon-gluon mixture is rather small in the astrophysical context under consideration.

For a degenerate MSQM, the energy density and pressures are given by the expressions~\cite{Felipe:2007vb}
\begin{equation}
\varepsilon=B\sum_i{\Omega}^0_{i}\sum_{\eta=\pm 1}\sum_{n}\left (
x_ip^{\eta}_{F,i}+h_{i}^{\eta\,\,2}\ln\frac{x_i+p_{F,i}^{\eta}}{h_{i}^{\eta}}\right
),\label{TQi}
\end{equation}
\begin{equation} \label{Ppar}
  P_{\parallel}=B\sum_i {\Omega}^0_{i}\sum_{\eta=\pm 1}\sum_n\left ( x_{i}p_{F,i}^{\eta} - h^{\eta\,\,2}_{i}\ln\frac{x_{i} +
 p^{\eta}_{F,i}}{h^{\eta}_{i}}\right ),\end{equation}
\begin{equation}\label{Pper}
 P_{\perp} = B\sum_i {\Omega}^0_{i}\sum_{\eta=\pm 1} \sum_{n}\left (2h_{i}^{\eta}\gamma_i^{\eta} \ln\frac{x_{i} + p_{F,i}^{\eta}}{h_{i}^{\eta}}\right),
\end{equation}
with
\begin{equation}
\gamma^{\eta}_i=\frac{B\,(2n+1-\eta)}{2B^c_i\sqrt{(2n+1-\eta)B/B^c_i+1}}-\eta y_i B\,.
\end{equation}
The number density is $N=\sum_i N_i$ with
\begin{equation}
 N_i = N^0_{i}\frac{B}{B^c_{i}}\sum_{\eta=\pm 1} \sum_{n} p^{\eta}_{F,i}\,, \quad
N_{i}^0 =  \frac{d_i m_{i}^3}{2\pi^2}\,.\label{TQf}
\end{equation}
The different expressions for the parallel pressure $P_{\parallel}$ and the transverse pressure $P_{\perp}$ reflect the anisotropy of pressure due to the magnetic field~\cite{Felipe:2008cm}. Notice however that for magnetic field values $B \lesssim 10^{19}$~G such an anisotropy is small.

In the astrophysical context, $\beta$-equilibrium is realized and the charge neutrality condition should be imposed. On the other hand, if one assumes that neutrinos enter and leave the star freely~\cite{Alford:2002kj} lepton number is not conserved inside the stars. For SQM, weak equilibrium relates chemical potentials as $\mu_u=\mu_B-\frac{2}{3}\mu_e$ and $\mu_d=\mu_s=\mu_B+\frac{1}{3}\mu_e$ and charge neutrality implies that
\begin{align}
2N_u-N_d-N_s=3N_e.
\end{align}

For the CFL phase, we shall write the equation of state using as starting point the fictional state of MSQM in which the baryon density $n_B$ equals each of the quark densities. The latter condition appears as a consequence of the minimization of the energy and the imposition of color and electric neutrality of the CFL phase~\cite{Alford:2001zr,Rajagopal:2000ff,Alford:2002kj,Alford:2007xm,Alford:2004pf}. There are no electrons in the CFL phase~\cite{Rajagopal:2000ff}.

We consider that the cost of the free energy which is compensated by the pairing formation is given as \begin{equation} \Omega_{\Delta}= \frac{3\Delta^2\mu_B^2}{\pi^2}, \end{equation} where $\Delta$ is the gap parameter. To simplify our study we do not consider any dependence of $\Delta$ on the magnetic field. Furthermore, we assume a common value of the gap parameter for the predominant color pairings ($ud,us,ds$). Such a gap pattern turns out to be a good approximation in the astrophysical scenario under consideration, namely, a compact quark star\footnote{In this case, typical densities are of the order of 500 MeV and for a magnetic field strength as large as $5\times 10^{18}$~G, one has $\sqrt{eB} \sim 172$~MeV, which implies $eB/\mu_B^2 < 1$~\cite{Fukushima:2007fc,Noronha:2007wg}.}. The study of how this gap is generated is out of the scope of the present work. Nevertheless, it is worth emphasizing that, even though strange quark matter is described in our framework as non-interacting, the presence of the gap is a consequence of the interaction among the quarks via BCS pairing.

The thermodynamical potential of the magnetized CFL phase is then written as
\begin{equation}\label{CFL}
\Omega_{CFL}= \Omega_{MSQM} - \Omega_{\Delta},
\end{equation}
and the energy density, derived from it, reads as
\begin{equation}\label{densityT}
\varepsilon_{CFL}=\Omega_{CFL}+3\mu_B\,n_B\,.
\end{equation}

Requiring the baryon density $n_B$ to equal each of the quark densities, the equations
\begin{equation}\label{Nequality}
N_u+\frac{2\Delta^2\mu_B}{\pi^2}=N_d+\frac{2\Delta^2\mu_B}{\pi^2} =N_s+\frac{2\Delta^2\mu_B}{\pi^2}=n_B
\end{equation}
are obtained. Due to the contribution of the magnetic masses, eqs.~\eqref{Nequality} must be solved
numerically. For illustration, in fig.~\ref{chempot} we present the dependence of the chemical potentials $\mu_B, \mu_3$ and $\mu_8$ on the baryon density $n_B/n_0$ ($n_0 \simeq 0.16$~fm$^{-3}$), obtained by solving eqs.~\eqref{Nequality} with $\Delta=50$~MeV and $B_{\rm bag}=75$~MeV/fm$^3$. The curves are given for two values of the magnetic field, $B=0$ (solid lines) and $B= 5\times 10^{18}$~G (dash-dotted lines). As can be seen from the figure, over the whole range of baryon density values $\mu_3 \simeq 0$, while $\mu_8$ is well approximated by the expression $\mu_8 \simeq -m_s^2/(2 \mu_B)$. These results are easily understood if one recalls that, in the absence of a magnetic field and neglecting the up and down quark masses, the neutrality conditions $\partial\Omega_{CFL}/\partial\mu_3=\partial\Omega_{CFL}/\partial\mu_8=0$ imply $\mu_3 = \mu_e$ and $\mu_8 = \mu_e/2 -m_s^2/(2 \mu_B)$~\cite{Alford:2002kj}.

\section{ Stability of CFL magnetized strange matter and EoS}
\label{sec3}

In this section we study the EoS and the stability of the magnetized CFL phase
within the MIT bag model. Our analysis is done varying the relevant input parameters of the model, i.e. the baryon density $n_B$, the magnetic field $B$, the bag $B_{\rm bag}$ and gap $\Delta$ as well as the strange quark mass $m_s$. Since in a strong magnetic field the anisotropy of pressures implies $P_{\perp} < P_{\parallel}$~\cite{Felipe:2007vb}, within the MIT bag framework the stability condition for quark matter is
\begin{equation} \label{stabpper}
P  =\sum_{i}P_{\perp,i}- B_{\rm bag}=0\,.
\end{equation}

\begin{figure}[t]
\begin{center}
\includegraphics[width=8.0cm]{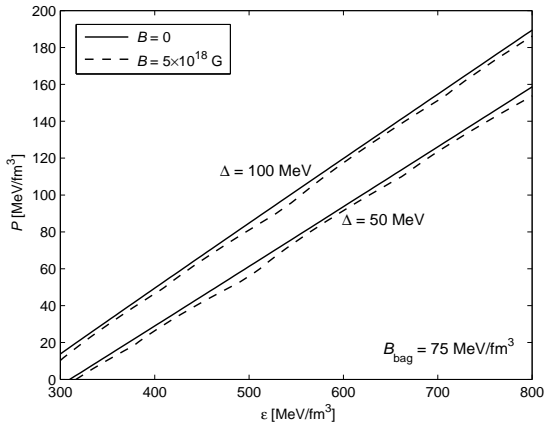}
\caption{EoS (pressure vs. energy) of the CFL phase ($B=0$) and magnetized CFL phase with $B=5\times10^8$~G. The curves are shown for two different values of the gap parameter, $\Delta= 50$~MeV and 100~MeV.\label{EoS}}
\end{center}
\end{figure}

\begin{figure*}[t]
      \begin{minipage}[t]{0.45\linewidth}
      \includegraphics[width=8.0cm]{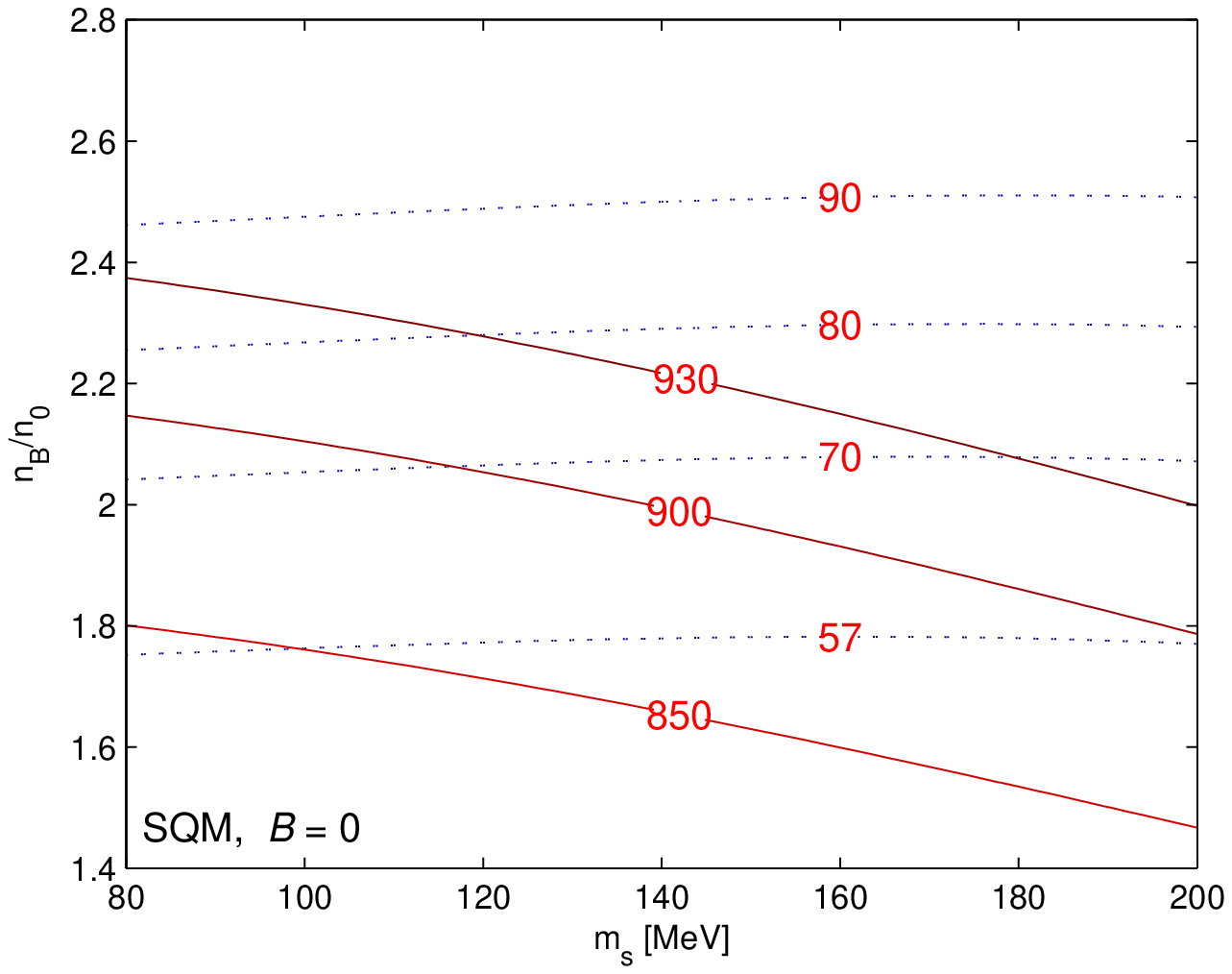}
      \end{minipage}
      \hspace{0.3cm}
      \begin{minipage}[t]{0.45\linewidth}
      \includegraphics[width=8.0cm]{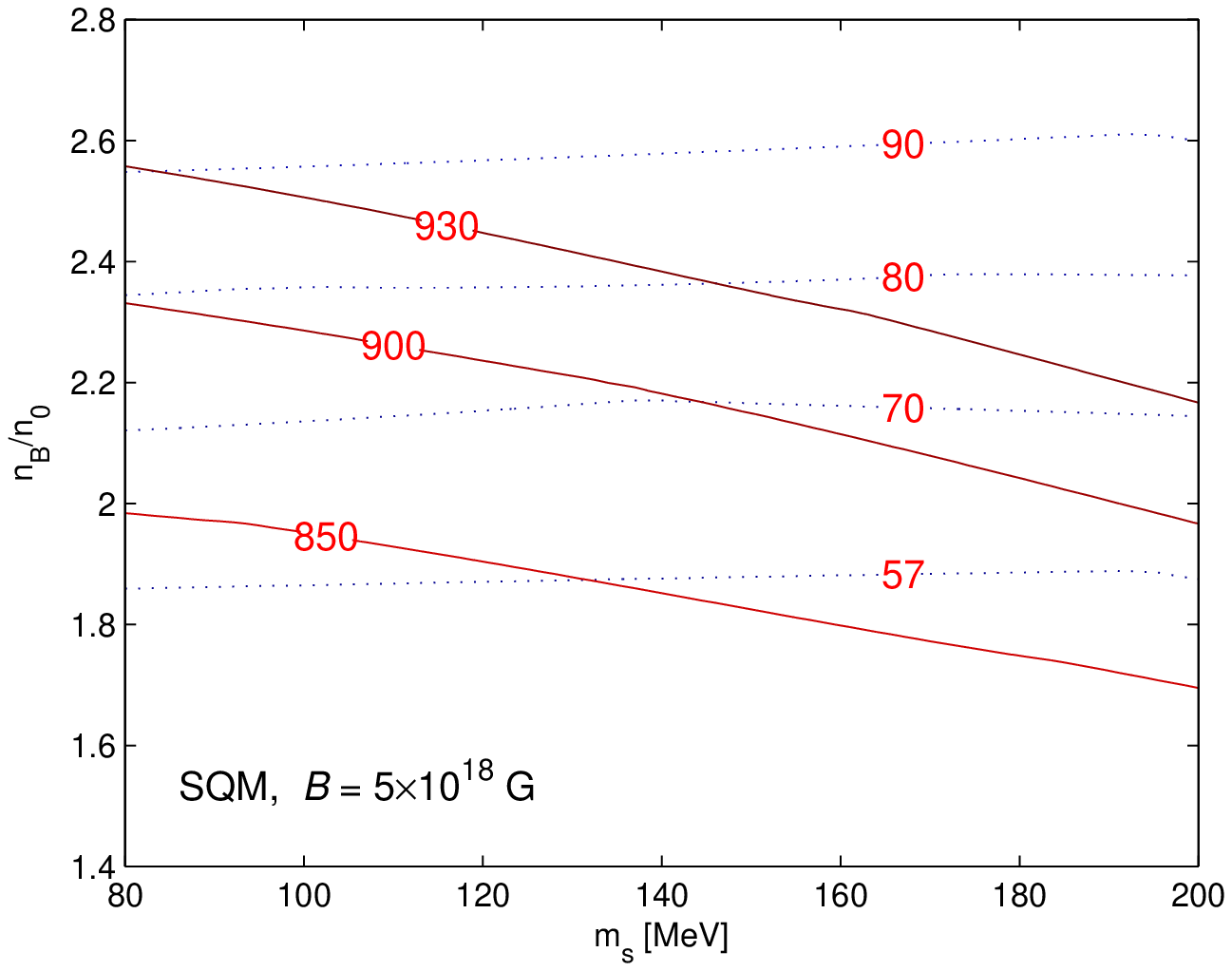}
      \end{minipage}
       \caption{Stability regions in the ($m_s,n_B$)-plane for SQM in the absence of a magnetic field and for magnetized SQM with $B=5\times 10^{18}$~G. The solid lines correspond to contours of the constant $E/A$ while the dashed lines represent the contours of the constant $B_{\rm bag}$.\label{contours1}}
\end{figure*}

In order to obtain the EoS of the magnetized CFL phase we solve the system of equations \eqref{Nequality} together with the condition \eqref{stabpper}. This allows us to obtain the parameter region which verifies the stability inequalities
\begin{equation}\label{stabineq}
\left.\frac{E}{A}\right|_{CFL}^B<\left.\frac{E}{A}\right|_{CFL}^{B=0} <\left.\frac{E}{A}\right|_{MSQM}<\left.\frac{E}{A}\right|_{^{56}\text{Fe}}
< \left.\frac{E}{A}\right|_{u,d}\,,
\end{equation}
where $\left. E/A\right|_{^{56}\text{Fe}} \simeq 930$~MeV is the energy per baryon of the iron nuclei and stability of any phase means that its energy per baryon is lower than this value. On the other hand, requiring the energy per baryon of normal matter (quark matter composed by $u$ and $d$ quarks) to be higher than the one of nuclear matter, i.e. $\left.E/A \right|_{u,d}> m_n$, where $m_n \simeq 939$~MeV is the neutron mass, yields the lower bound $B_{\rm bag}> 57$~MeV/fm$^3$~\cite{Farhi:1984qu} under the same conditions at $P=0$ and $T=0$.

In fig.~\ref{EN} we present a comparison of the energy per baryon $E/A$ as a function of the baryon density $n_B/n_0$ in the absence of a magnetic field (left plot) and for a magnetic-field value of $5\times 10^{18}$~G (right plot). Both plots are given for $B_{\rm bag}=75$~MeV/fm$^3$ and two different values of the gap, $\Delta=50,100$~MeV. The corresponding curves for the SQM ($B=0$) and MSQM ($B\neq 0$) states are also included for comparison. As can be seen from the figure, for a fixed value of the baryon density, the energy per baryon of the magnetized CFL phase is lower than its corresponding value in the absence of a magnetic field.

The behavior of $E/A$ with the (transverse) pressure $P$ is shown in fig.~\ref{EP}. One can notice that the point of zero pressure for the magnetized CFL phase is reached at an energy density value lower than for CFL without magnetic field. Consequently, magnetized CFL matter is more stable and more bound. A comparison of the baryon density at the zero-pressure point for the CFL state without magnetic field and with $B= 5\times 10^{18}$~G is presented in table~\ref{table1}. From the table we conclude that in the presence of the magnetic field the zero-pressure point is attained at slightly higher values of the baryon density.

\begin{table}
\caption{Comparison of the baryon density at the zero-pressure point for the CFL state without magnetic field and in the presence of a strong magnetic field. We assume $B_{\rm bag}=75$~MeV/fm$^3$.}
\label{table1}
\begin{center}
\begin{tabular}{ccc}
$\Delta$~(MeV)& $B$~(G) & $n_B/n_0$ \\\hline\hline
50 & 0 & $2.15$\\
 & $5\times 10^{18}$ & $2.21$\\\hline
100 & 0 & $1.96$ \\
 & $5\times 10^{18}$ & $2.08$\\\hline
\end{tabular}
\end{center}
\end{table}

To determine the EoS of the magnetized CFL phase, the system of equilibrium conditions must be solved numerically. In fig.~\ref{EoS} we present the EoS for the CFL phase (i.e. when $B=0$) and for the magnetized CFL state at $B=5\times 10^{18}$~G, for two different values of the gap parameter, $\Delta= 50$ and 100~MeV. As can be seen from the figure, the EoS of a strongly magnetized CFL phase is softer than the EoS of the CFL phase without a magnetic field (i.e. it produces less pressure for a given energy density). This in turn implies that macroscopic observables, such as the star mass and radius, will be modified. In particular, since a soft EoS can sustain less gravitational force, it will lead to compact stars with smaller maximum mass values (cf. table~\ref{table2} in the next section).

\begin{figure*}[t]
      \begin{minipage}[t]{0.45\linewidth}
      \includegraphics[width=8.0cm]{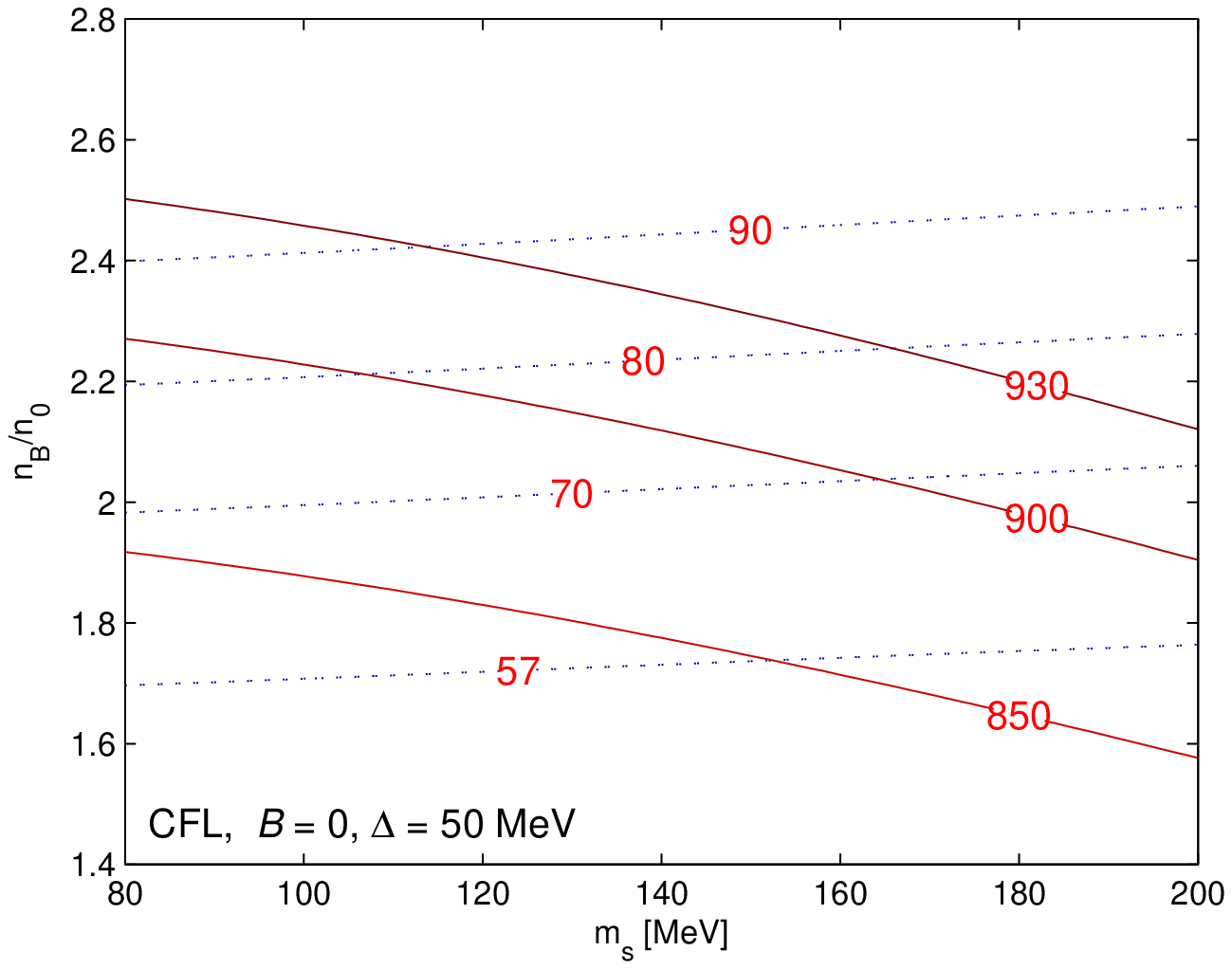}
      \end{minipage}
      \hspace{0.3cm}
      \begin{minipage}[t]{0.45\linewidth}
      \includegraphics[width=8.0cm]{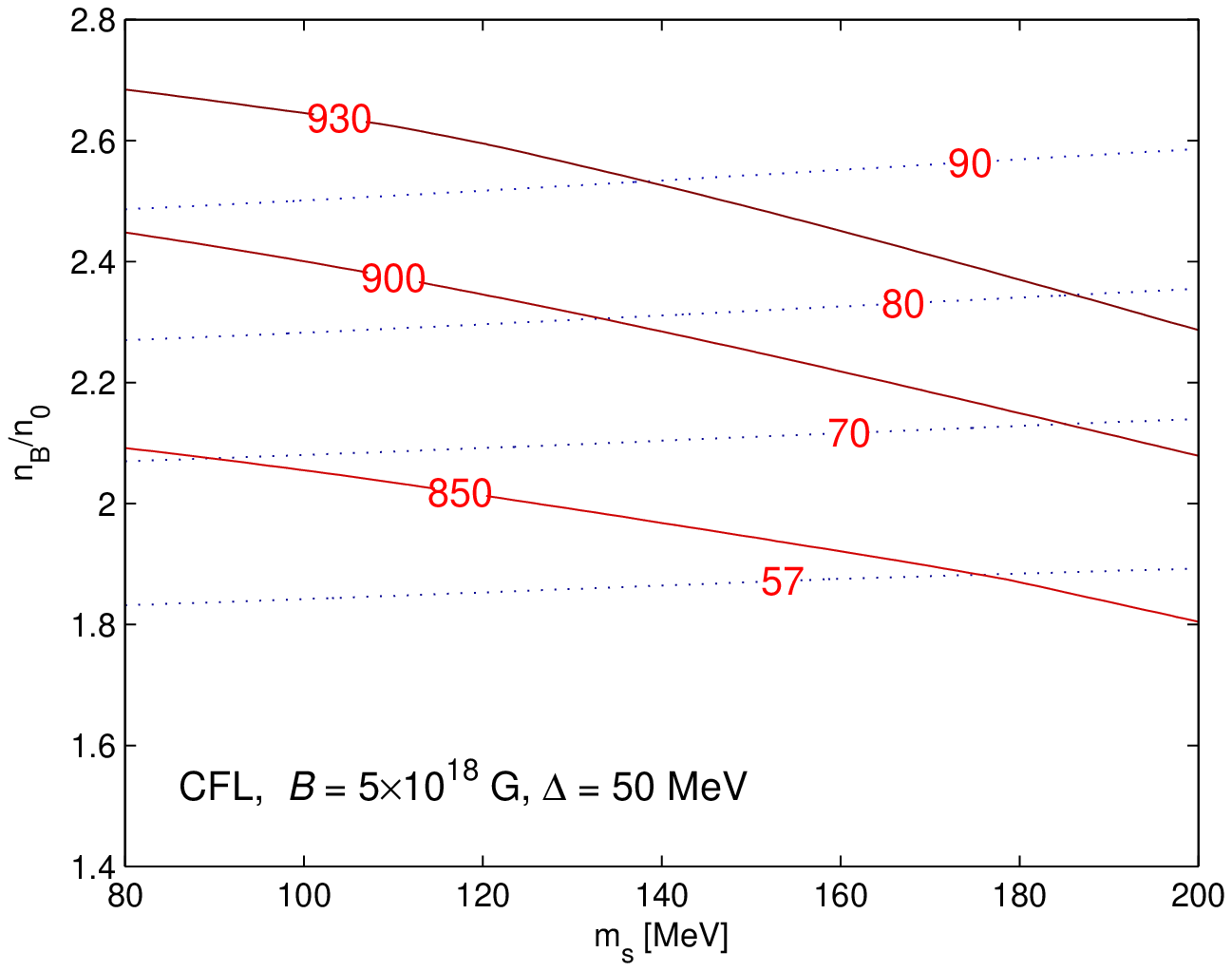}
      \end{minipage}\\
      \begin{minipage}[t]{0.45\linewidth}
      \includegraphics[width=8.0cm]{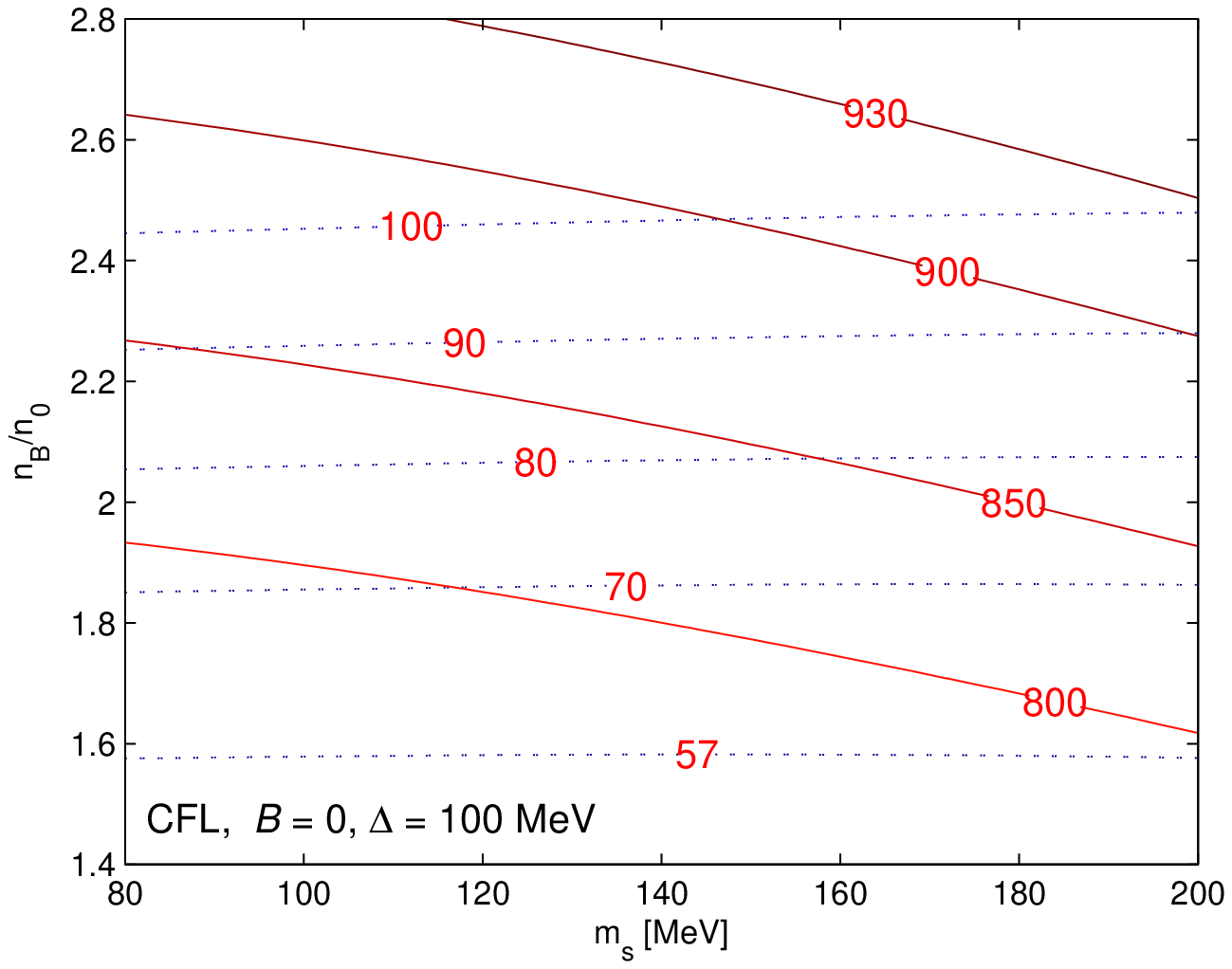}
      \end{minipage}
      \hspace{0.3cm}
      \begin{minipage}[t]{0.45\linewidth}
      \includegraphics[width=8.0cm]{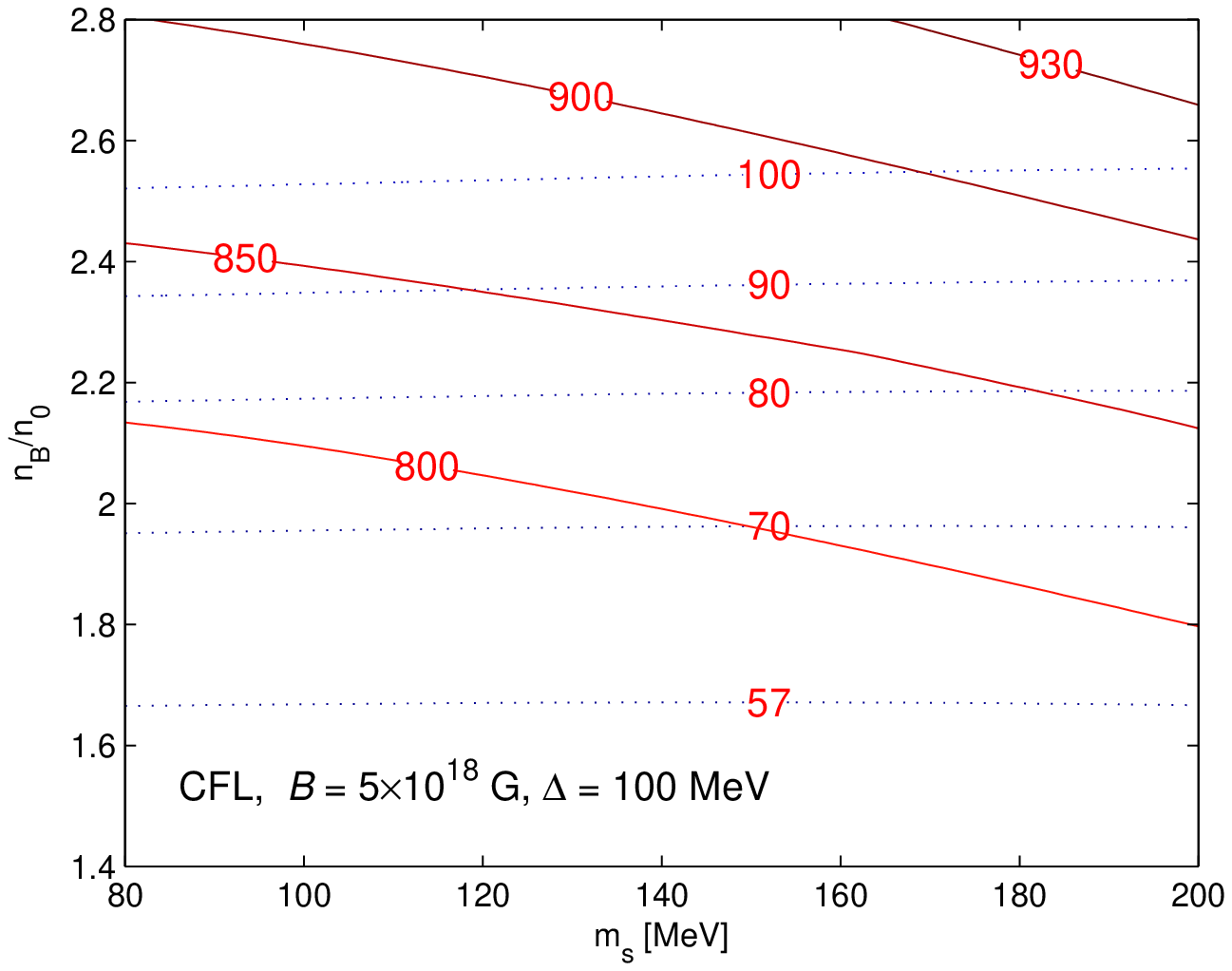}
      \end{minipage}
       \caption{As in Figure~\ref{contours1}, but for the CFL phase. We take $\Delta=50$ and $100$~MeV.\label{contours2}}
\end{figure*}

Next we study the stability window for the magnetized CFL phase, i.e. the allowed region of the baryon density, strange quark mass and bag and gap parameters for a given magnetic field value. In order to investigate how the magnetic field affects this window, we consider first the stability regions in the ($m_s,n_B$)-plane for fixed values of the magnetic field and the gap parameter. In fig.~\ref{contours1} we present the contours of $B_{\rm bag}$ and $E/A$ of the SQM (left plot) and magnetized SQM phase (right plot) for a magnetic field value of $5\times 10^{18}$~G. The corresponding contours for the CFL phase are shown in fig.~\ref{contours2}. All the contours were obtained by imposing eq.~\eqref{stabpper}, i.e. the vanishing of the pressure. For the CFL case we have fixed $\Delta=50$ and $100$~MeV. In both cases, it is evident from the figures that increasing the value of the gap parameter shifts the curves of the constant $E/A$ to higher values of the baryon density, while the contours of constant $B_{\rm bag}$ are displaced to lower values of $n_B$.

\begin{figure*}[t]
      \begin{minipage}[t]{0.45\linewidth}
      \includegraphics[width=8.0cm]{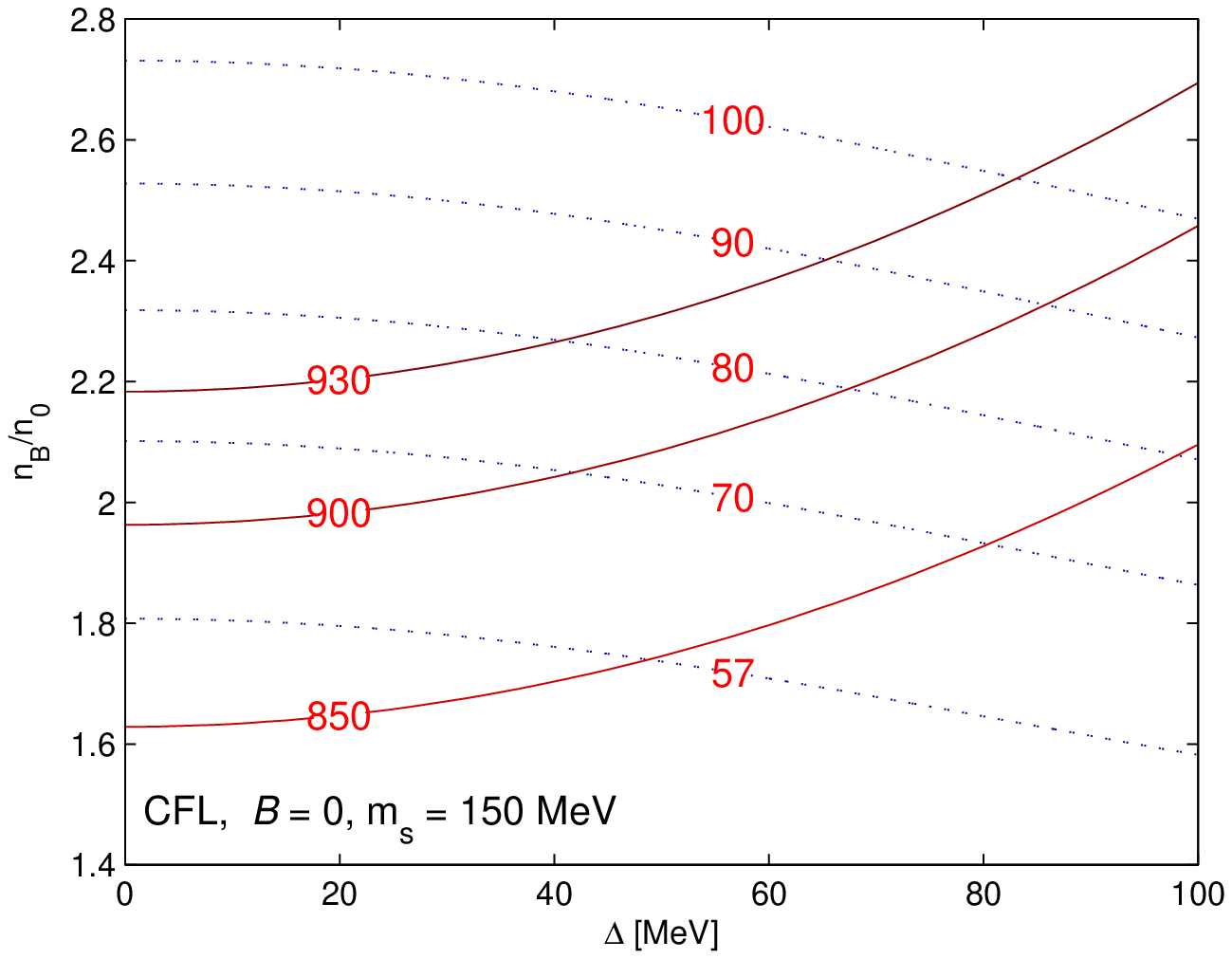}
      \end{minipage}
      \hspace{0.3cm}
      \begin{minipage}[t]{0.45\linewidth}
      \includegraphics[width=8.0cm]{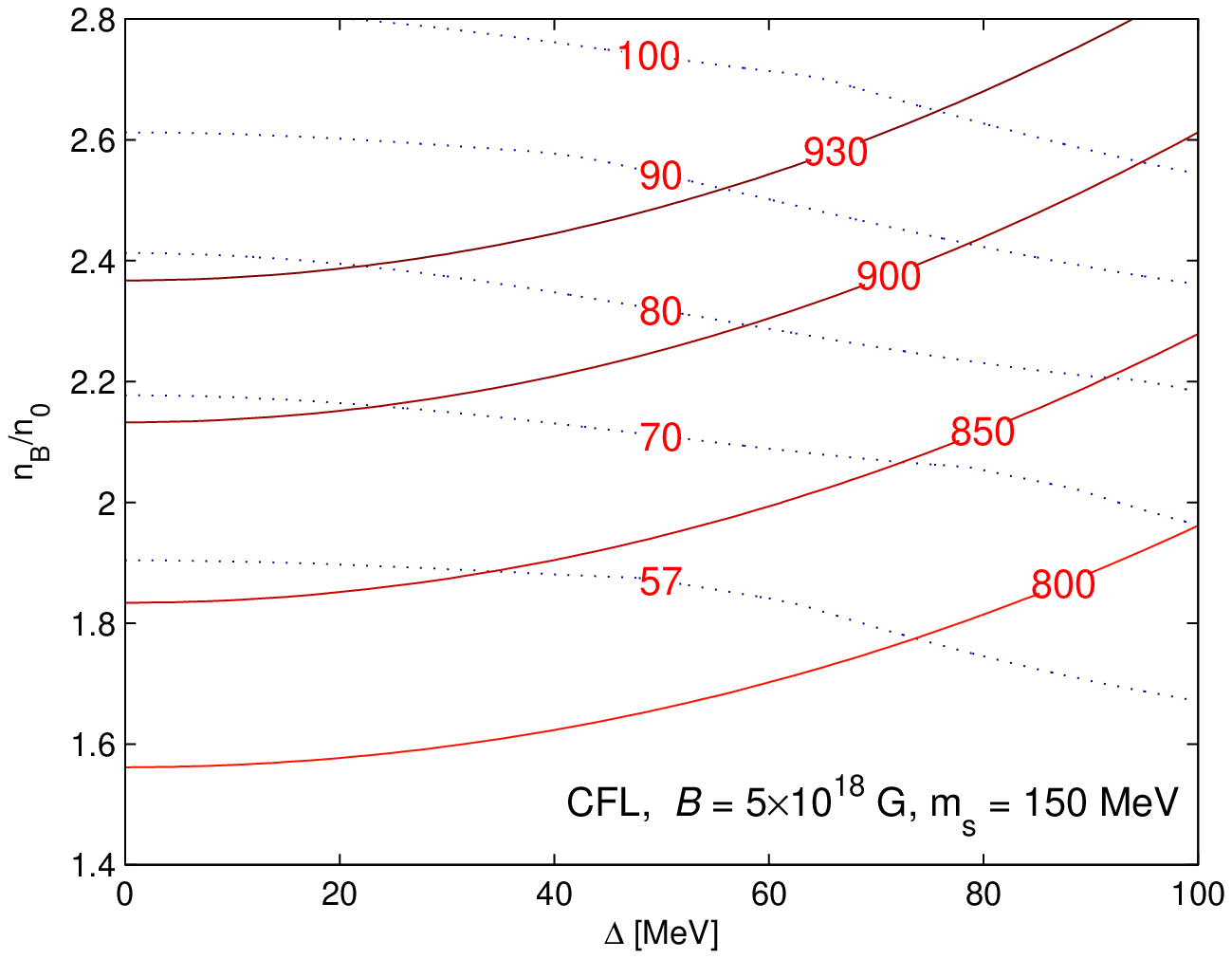}
      \end{minipage}
      \caption{Contours of $E/A$ and $B_{\rm bag}$ in the ($\Delta,n_B$)-plane for the CFL state with $B=0$ (left plot) and $B=5\times 10^{18}$~G (right plot) and a fixed value of the strange quark mass, $m_s = 150$~MeV.\label{contours3}}
\end{figure*}

Finally, in fig.~\ref{contours3} we present the contours of $E/A$ and $B_{\rm bag}$ in the ($\Delta,n_B$)-plane for the CFL state with $B=0$ (left plot) and $B=5\times 10^{18}$~G (right plot) and a fixed value of the strange quark mass, $m_s = 150$~MeV. Note that, as the gap parameter $\Delta$ increases, the stability windows (i.e. the allowed range of the baryon density and bag parameter for a given energy per baryon) is enlarged.

\section{Magnetized CFL state and mass-radius relation}
\label{sec4}

\begin{figure*}[t]
      \begin{minipage}[t]{0.45\linewidth}
      \includegraphics[width=8.0cm]{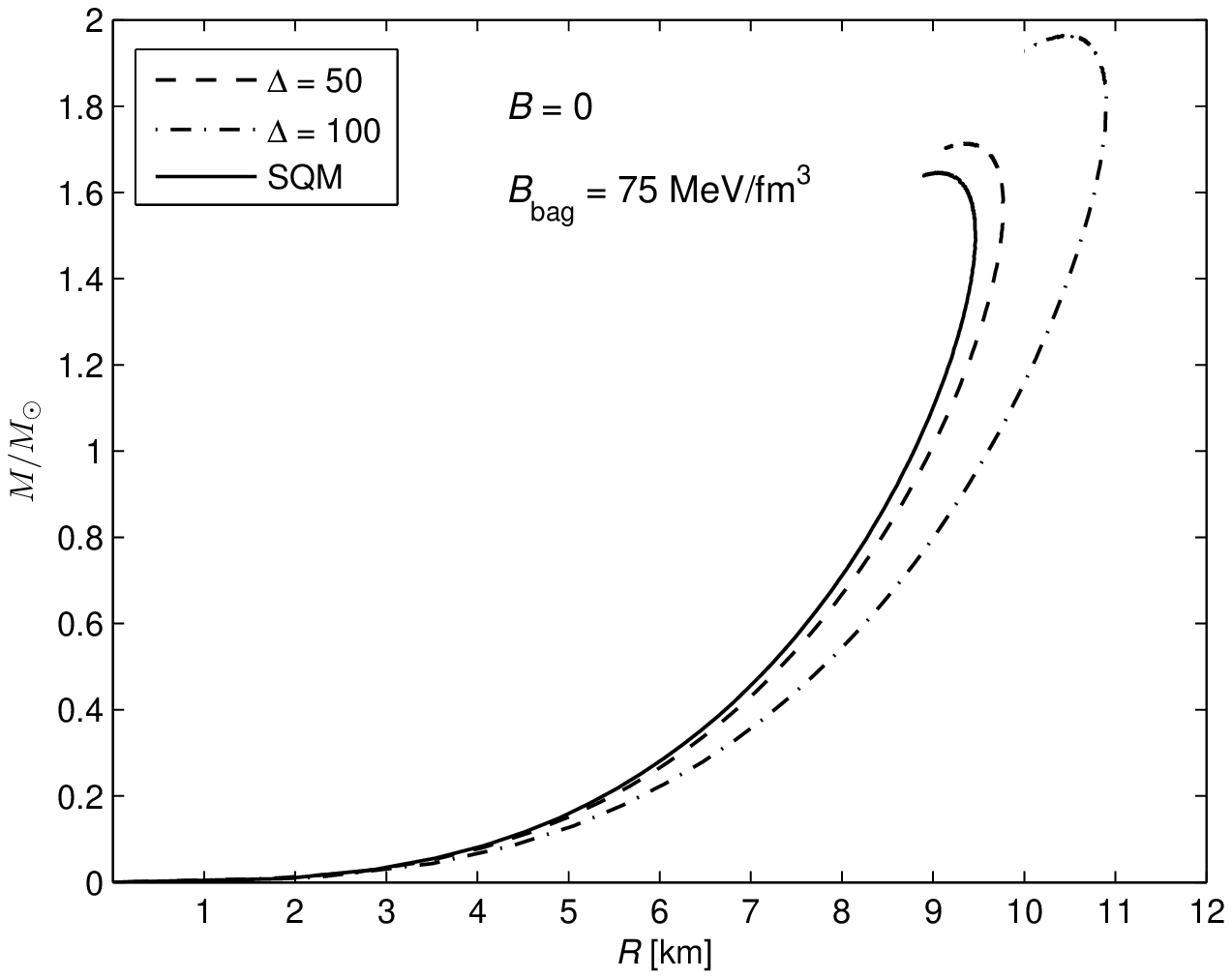}
      \end{minipage}
      \hspace{0.3cm}
      \begin{minipage}[t]{0.45\linewidth}
      \includegraphics[width=8.0cm]{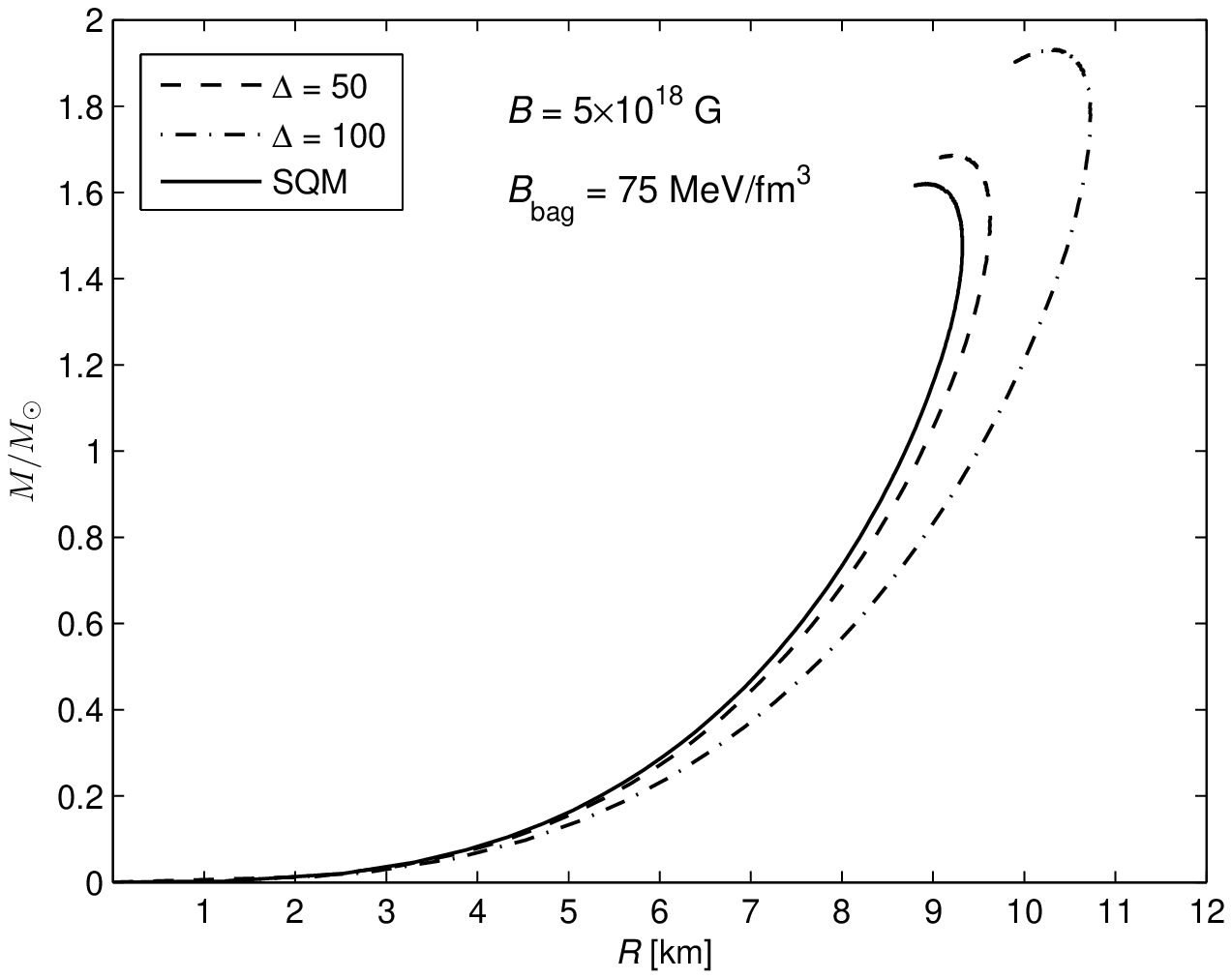}
      \end{minipage}
\caption{Stable $M$-$R$ configurations for CFL strange stars in the absence of a magnetic field (left plot) and for $B=5\times 10^{18}$~G (right plot). The curves are given for two different values of the gap parameter: $\Delta=50,100$~MeV. The solid lines correspond to the standard SQM phase.\label{tov}}
\end{figure*}

Let us now study the equilibrium configuration of magnetized strange quark stars described by the EoS of the CFL phase obtained in the previous section. As is well known, the most important macroscopic parameters of a star are its radius $R$ and its gravitational mass $M$. Configurations of spherical symmetric non-rotating compact stars are obtained by the numerical integration of the TOV equations~\cite{Baym},
\begin{eqnarray}
\frac{dM}{dr}&=&4\pi r^2\varepsilon (r)\,,\nonumber\\
\frac{dP}{dr}&=& -G\frac{\left(\varepsilon (r) +P (r)\right)\left(
M(r) + 4\pi P(r) r^3 \right)}{r^2-2 G r M (r)}\,, \label{TOV}
\end{eqnarray}
supplemented with the EoS, where $P(r)$ is the pressure and $\varepsilon(r)$ is the energy density. The radius $R$ and the corresponding mass $M$ of the star are determined by the value of $r$ for which the pressure vanishes, $P(R)=0$. The EoS fixes the central pressure, which together with the condition $M(0)=0$, completely determine the system of equations \eqref{TOV}. In this way, varying continuously the central pressure, one obtains a mass-radius relation $M(R)$, which relates masses and radii for a given EoS. We also note that the magnetic field contributes to the total pressure and energy of the star through the magnetic field pressure and energy density term $B^2/(8\pi)$. Nevertheless, the latter could be viewed in our framework as a field-dependent redefinition of the bag parameter and, for the magnetic field values considered, can be safely neglected.

In fig.~\ref{tov} we plot the stable $M$-$R$ configurations of CFL strange stars for $B=0$ and $B=5\times 10^{18}$~G. For comparison, the usual SQM phase is also depicted (solid lines). The curves are presented for two values of the gap parameter: $\Delta=50$ and 100~MeV. The bag parameter has been chosen as 75~MeV/fm$^3$. The corresponding maximum masses, $M_{\rm max}$, and maximum radii, $R_{\rm max}$, of magnetized CFL strange stars are summarized in tables~\ref{table2} and \ref{table3}. As expected, the effect of increasing the magnetic field, while holding the gap $\Delta$ and bag $B_\text{bag}$ fixed, is to yield more compact stars (with smaller masses and radii). Note also that an increase of $\Delta$ leads to higher values of both $M_{\rm max}$ and $R_{\rm max}$. Nevertheless, for magnetic fields around $5\times 10^{18}-10^{19}$~G and fixed values of the bag and gap parameters, the values of $M_{\rm max}$ obtained turn out to be consistent with the measurements of maximum masses~\cite{Cottam:2002cu}.

\begin{table*}[t]
\caption{Maximum mass $M_\text{max}$ and the corresponding radius for different values of the
magnetic field and gap parameter.}\label{table2}
\begin{center}
\begin{tabular}{ccccccc}
& SQM &    & $\Delta=50$ MeV &  & $\Delta=100$
MeV \\\hline\hline B (G) & $M_\text{max}/M_{\odot}$ & $R$ (km) &
$M_\text{max}/M_{\odot}$ & $R$
(km)& $M_\text{max}/M_{\odot}$ & $R$ (km)\\ \hline 0 & 1.65 & 9.07 & 1.71 & 9.40 & 1.96 & 10.48\\ \hline $5\times 10^{18}$ & 1.62 & 8.91 & 1.69 & 9.25 & 1.93 & 10.30\\ \hline
$10^{19}$ & 1.56 & 8.62 & 1.62 & 8.93 & 1.86 & 9.93 \\\hline
\end{tabular}
\end{center}
\end{table*}

\begin{table*}[t]
\caption{Maximum radius $R_\text{max}$ and the corresponding mass for different values of the
magnetic field and gap parameter.}\label{table3}
\begin{center}
\begin{tabular}{ccccccc}
& SQM &    & $\Delta=50$ MeV &  & $\Delta=100$
MeV \\\hline\hline B (G) &$R_\text{max}$ (km)  & $M/M_{\odot}$
& $R_\text{max}$ (km)
&$M/M_{\odot}$ &$R_\text{max}$ (km)  & $M/M_{\odot}$\\ \hline  0 &
9.47 & 1.49 & 9.77 & 1.56 & 10.90 & 1.82\\ \hline
$5\times 10^{18}$ & 9.32 & 1.50 & 9.63 & 1.54 & 10.73
& 1.78\\ \hline $10^{19}$ & 9.00 & 1.41 & 9.30 & 1.49 & 10.39 & 1.70 \\ \hline
\end{tabular}
\end{center}
\end{table*}

\section{Conclusions}
\label{sec5}

In the present work we have investigated the stability of the magnetized CFL phase within the phenomenological MIT bag model. The study was performed taking into account the variation of the strange quark mass, the baryon density, the magnetic field, the gap and the bag parameters. We found that a strongly magnetized CFL state of strange matter is indeed more stable than the non-magnetized CFL one. We have also shown that magnetized CFL matter is more stable than unpaired magnetized SQM in the range of strong magnetic fields typically expected in compact stars and for a wide range of values of the gap of the QCD Cooper pairs. As the pairing gap increases, the stability windows of the CFL phase is enlarged, but at the same time the EoS becomes stiffer and, consequently, the maximum mass and radii values of stable stellar configurations get larger~\cite{Lugones:2002zd}. On the other hand, an increase of the bag parameter $B_\text{bag}$ would lead to smaller values of $M_\text{max}$ and $R_\text{max}$.

Although our results have been obtained in a simplified framework of the MIT bag model, it is worth pointing out that relaxing our simplifying assumptions would not significantly change our qualitative conclusions. As illustrated in fig.~\ref{tov} and tables~\ref{table2} and~\ref{table3}, the derived mass-radius relation and the relevant macroscopic parameters of compact stars are indeed modified in the presence of a strong magnetic field. In particular, the derived EoS for the magnetized CFL phase turns out to be softer (when compared to the EoS in the absence of a magnetic field), thus allowing for the existence of very compact stable configurations of strange quark stars composed by deconfined matter in a CFL phase. The magnetic field in these compact objects will be frozen over a time scale comparable with the age of our Universe. This feature enriches their phenomenology and distinguishes canonical neutron stars from those with color superconducting quark cores. The variation of the masses and radii between the non-magnetized and magnetized CFL phases are nevertheless rather small. Thus, it seems difficult to prove the existence of color superconductivity of magnetized quark matter by measurements of the $M(R)$ relation alone. More promising approaches are those based on the effects of the magnetic field in transport properties, such as neutrino emission~\cite{Reddy:2002xc}, bulk and shear viscosity~\cite{Madsen:1999ci} and glitches~\cite{Alford:2000ze}.

There are several ways in which our study could be extended further. In our analysis we did not take into account perturbative QCD corrections to the equation of state and we have neglected a possible dependence of the gap parameter on the density and the magnetic field. It would be important to see how robust our conclusions are against these and other possible corrections. Finally, it would also be interesting to extend our study to hybrid compact stars with a nuclear matter crust and a quark matter core in the presence of a strong magnetic field, pursuing for instance a model-independent phenomenological approach as the one taken in ref.~\cite{Alford:2004pf}.

\begin{acknowledgement}
We are grateful to J. Horvath, E.~J.~Ferrer and V.~de la Incera for very useful comments and suggestions. The work of R.G.F. was supported by \emph{Funda\c{c}\~{a}o para a Ci\^{e}ncia e a Tecnologia} (FCT, Portugal) through the project CFTP-FCT UNIT 777, which is partially funded through POCTI (FEDER). The work of A.P.M. has been supported by \emph{Ministerio de Ciencia, Tecnolog\'{\i}a y Medio Ambiente} under the grant CB0407 and the ICTP Office of External Activities through NET-35. A.P.M. also acknowledges TWAS-UNESCO PCI Programme for financial support at CBPF-Brazil.
\end{acknowledgement}

\end{document}